\documentclass[journal]{IEEEtran}
\IEEEoverridecommandlockouts
\usepackage{cite}
\usepackage{hyperref}
\usepackage{amsmath,amssymb,amsfonts}
\usepackage{amsmath}
\usepackage{amsthm}

\usepackage{graphicx}
\usepackage{textcomp}
\usepackage{xcolor}
\usepackage{graphicx}
\usepackage{float}
\usepackage{subfigure}
\usepackage{amsmath}
\usepackage{amsfonts,amssymb}
\usepackage{mathrsfs}
\usepackage{mathtools}
\usepackage{algorithm}
\usepackage{algorithmicx}
\usepackage{algpseudocode}
\usepackage{siunitx}
\usepackage{bm}
\makeatletter
\newcommand{\biggg}{\bBigg@{3}}
\newcommand{\Biggg}{\bBigg@{3.5}}
\makeatother
\usepackage{stfloats}
\newcounter{TempEqCnt}
\bibliography{IEEEabrv}
\def\BibTeX{{\rm B\kern-.05em{\sc i\kern-.025em b}\kern-.08em
    T\kern-.1667em\lower.7ex\hbox{E}\kern-.125emX}}
\newtheorem{lemma}{Lemma}
\begin{document}

\title{Secrecy Performance for Finite-Alphabet Inputs Over Fluctuating Two-Ray Channels in FDA Communications\\
}

\author{Chongjun~Ouyang,
        Sheng~Wu,
        Chunxiao~Jiang,~\IEEEmembership{Senior Member,~IEEE,}\\
        Derrick~Wing~Kwan~Ng,~\IEEEmembership{Senior Member,~IEEE,}
        and~Hongwen~Yang
\thanks{The work of C. Jiang was supported by the National Natural Science Foundation of China (No. 61922050). C. Ouyang is supported by BUPT Excellent Ph.D. Students Foundation (CX2020101). D. W. K. Ng is supported by funding from the UNSW Digital Grid Futures Institute, UNSW, Sydney, under a cross-disciplinary fund scheme and by the Australian Research Council's Discovery Project (DP190101363).}
\thanks{C. Ouyang, S. Wu and H. Yang are with the School of Information and Communication Engineering, Beijing University of Posts and Telecommunications, Beijing, 100876, China. E-mail: \{DragonAim, thuraya, yanghong\}@bupt.edu.cn.}
\thanks{C. Jiang is with the Tsinghua Space Center, Tsinghua University, Beijing 100084, China, and also with the Beijing National Research Center for Information Science and Technology, Tsinghua University, Beijing 100084, China. E-mail: jchx@tsinghua.edu.cn.}
\thanks{D. W. K. Ng is with the School of Electrical Engineering and Telecommunications, the University of New South Wales, Sydney, NSW 2052, Australia. E-mail: w.k.ng@unsw.edu.au.}
}

\maketitle

\begin{abstract}
To provide system design insights for practical communication systems equipped with the frequency diverse array (FDA), this paper investigates the secrecy performance of FDA systems exploiting finite-alphabet inputs over fluctuating two-ray (FTR) fading channels. More specifically, closed-form expressions for the average secrecy rate (ASR) and the secrecy outage probability (SOP) are derived, while their correctness is confirmed by numerical simulations. In addition, we perform asymptotic analysis to quantify the secrecy performance gap between Gaussian and finite-alphabet inputs, for a sufficiently large average signal-to-noise ratio (SNR) of the main channel. Compared with Gaussian inputs-based research, this letter focuses on practical scenarios which sheds lights on properties of FDA systems.
\end{abstract}


\section{Introduction}
As a charming aspect of wireless networks, physical layer security (PLS) has aroused considerable attention from both the academy and the industry \cite{b17}. In the field of PLS, transmit beamforming (TB) is an important research topic and frequency diverse array (FDA) is a typical hardware architecture for realizing practical TB \cite{b2}. By properly setting the frequency offset of each array element in an FDA, a joint-range-direction-dimension beampattern can be realized in contrast to the direction-dimension-only beampattern realized by a conventional phased array (PA) \cite{b3}. Thus, PA systems can only achieve secure communications in the direction dimension, whereas FDA systems can achieve secure high-data rate transmission in both range dimension and direction dimension  \cite{b2,b18,b3,b4,b5}.

On a parallel avenue, conducting secrecy performance analysis is fundamental for system engineers. By deriving the exact or approximated expressions to characterize the average secrecy rate (ASR) and the secrecy outage probability (SOP), important system insights can be revealed and further be exploited to improve the system design. As a pioneering technology to guarantee PLS, FDA has gained much research interest and its secrecy performance has been widely discussed \cite{b4,b5,b18}. For instance, Hu {\emph{et al.}} derived an analytical upper bound for secrecy rate in FDA systems over additive white Gaussian noise (AWGN) channels \cite{b18}. Later, \cite{b4,b5} extended the analysis in \cite{b18} to multi-path fading channels. Although the aforementioned works are fascinating, their derived results may be less referable for practical system design. In fact, these works assumed the channel inputs follow Gaussian distribution. Yet, as is well known, the input signals in modern wireless communication systems are drawn from a set of discrete finite alphabets, and thus they are non-Gaussian in general \cite{b12,b22}. Hence, there is a strong need to analyze the secrecy performance of FDA systems exploiting finite-alphabet inputs. However, the results available so far in the literature are greatly limited.

This study aims to investigate the secrecy performance of FDA systems adopting finite-alphabet inputs so as to fill the knowledge gap stated before. Specifically, a squared $M$-ary quadrature amplitude modulation ($M$-QAM) scheme is considered due to its extensive applications in practical systems. Most importantly, this work establishes a general analytical framework for evaluating the transmission security in wiretap fading channels having finite-alphabet inputs. Besides, we incorporate the newly developed fluctuating two-ray (FTR) fading model \cite{b8} into the analysis, which can capture a variety of physical environments and holds a wide generality. Besides, this model encompasses many classical statistical models as special scenarios, such as Rayleigh, Nakagami-$m$, and Rician fading \cite{b8}. {Note that the PLS over FTR fading channels was studied in \cite{b13,b24}, but the derived results therein were based on the assumption of Gaussian signaling instead of finite-alphabet signaling. For other more general fading channels such as the $\alpha$-$\eta$-$\kappa$-$\mu$ fading channel \cite{b29}, their secrecy performance can be also discussed following the analytical framework established in this paper, which will be left to our future work.}

To the best of authors' knowledge, our work is the first attempt to characterize the secrecy performance of FDA systems adopting discrete inputs, which offers novel and important system design insights. {Particularly, the unveiled system insights are summarized as follows: 1) in high signal-to-noise ratio (SNR) regimes of the main channel, the ASR and SOP achieved by finite-alphabet inputs converge to finite positive constants, and the rate of convergence is ${\mathcal O}\left(\bar\gamma_{\rm B}^{-1}\right)$, where $\bar\gamma_{\rm B}$ is the average SNR of the main channel; 2) FDA systems outperform the PA systems in terms of PLS in the range dimension, and the security achieved by an FDA can be enhanced by increasing the frequency offset or the number of antenna elements.}

\section{System Model}
\begin{figure}[!t]
\centering
\setlength{\abovecaptionskip}{-5pt}
\includegraphics[width=0.2\textwidth]{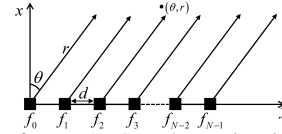}
\caption{The structure of an FDA where the receiver is located at $\left(r,\theta\right)$. Note that $f_{i}=f_{0}+i\Delta f$, $i=0,1,\cdots,N-1$.}
    \label{figure_FDA}
    \vspace{-15pt}
\end{figure}

Consider a multiple-input single-output single-antenna eavesdropper (MISOSE) wiretap fading channel, where a desired receiver (Bob) and an eavesdropper (Eve) are single-antenna devices. Moreover, a transmitter (Alice) is equipped with an FDA and its structure is shown in {\figurename} \ref{figure_FDA} \cite{b3}. For the receiver at coordinate $\left(r,\theta\right)$, the transmit steering vector adopted at Alice can be written as \cite{b4,b5}
\begin{equation}
{\bf v}\left(r,\theta\right)
=\frac{1}{\sqrt{N}}\left[1,{\rm e}^{\jmath 2\pi w\left(r,\theta\right)},\cdots,{\rm e}^{\jmath 2 \pi \left(N-1\right)w\left(r,\theta\right)}\right],\label{equation1}
\end{equation}
where $N$ is the number of antenna elements, $\jmath$ is the imaginary term, $w\left(r,\theta\right)=\left({f_0 d\sin\theta-r\Delta f}\right)/{c}$, $c$ is the light speed, $f_{0}$ is the carrier frequency, $\Delta f$ is the frequency offset, and $d$ is the inter-element space between antennas. Generally, $d$ is set to half of the wavelength, namely $d=\frac{c}{2f_0}$ \cite{b4,b5,b18}. Denote the coordinates of Bob and Eve as $\left(r_i,\theta_i\right)$, $i\in\{{{B}},{{E}}\}$, respectively. Based on \cite{b18,b4,b5}, the received signal of Bob or Eve can be written as $y_{i}={\sqrt{P}}{\sqrt{A_i}}h_i{\bf v}\left(r_i,\theta_i\right){\bf p}^{\dagger}s+n_{i}$, where $\left(\cdot\right)^{\dag}$ is the conjugate transpose operator, $P$ is the transmit power, $A_i=1/A\left(f_{0},r_i\right)$ is the free-space path loss, $n_{i}\sim{\mathcal{CN}}\left(0,\sigma_i^2\right)$ is the additive white Gaussian noise with power $\sigma_i^2$, $s$ is the normalized unit-power transmitted $M$-QAM symbol, $h_i$ is the FTR fading coefficient, and ${\bf p}\in{\mathbb C}^{1\times N}$ (${\mathbb C}$ denotes the complex plane) is the precoding vector adopted at Alice. To maximize the received SNR of Bob, Alice adopts the maximum ratio transmission (MRT) scheme, namely ${\bf p}={\bf v}\left(r_{\rm B},\theta_{\rm B}\right)$ \cite{b18,b4,b5}. Thus, the instantaneous received SNRs $\gamma_i$ of Bob and Eve can be written as $\gamma_i=\frac{\left|h_i\right|^2{\left|\Theta_i\right|^2}P}{\sigma_i^2A\left(f_{0},r_i\right)}$, where ${\Theta}_i\triangleq{\bf v}\left(r_i,\theta_i\right){\bf v}\left(r_{\rm B},\theta_{\rm B}\right)^{\dag}$. {We note that the MRT precoding is generally a low-complexity suboptimal scheme. In fact, the optimal precoding scheme for FDA systems having discrete inputs still remains as an open problem. Thus, the analysis of its secrecy performance will be considered in our future work.} Without loss of generality, we assume that the average small-scale fading power $\mathbb{E}\left\{\left|h_i\right|^2\right\}$ equals one and thus the average SNR is $\bar\gamma_i=\frac{{\left|\Theta_i\right|^2}P}{\sigma_i^2A\left(f_{0},r_i\right)}$. Under the FTR fading, the probability density functions (PDFs) and cumulative distribution functions (CDFs) of $\gamma_i$ are, respectively, given by \cite{b8}
\begin{align}
\label{eq1a}
f_i\left(\gamma_i\right)&=\frac{m_i^{m_i}}{\Gamma\left(m_i\right)}\sum_{j=0}^{\infty}\frac{K_i^jd^{\left(i\right)}_j\gamma_i^j}{j!j!u_i^{j+1}}\exp\left(-\frac{\gamma_i}{u_i}\right),\\
\label{eq1b}
F_i\left(\gamma_i\right)&=\frac{m_i^{m_i}}{\Gamma\left(m_i\right)}\sum_{j=0}^{\infty}\frac{K_i^jd^{\left(i\right)}_j}{j!j!}\Upsilon\left(j+1,\frac{\gamma_i}{u_i}\right),
\end{align}
where $d^{\left(i\right)}_j=\sum_{k=0}^{j}\binom{j}{k}\left(\frac{\Delta_i}{2}\right)^k\sum_{l=0}^{k}\binom{k}{l}\Gamma\left(j+m_i+2l-k\right)
\times{\rm e}^{\frac{\pi\left(2l-k\right)\jmath}{2}}\tau_i^{-\frac{j+m}{2}}
P_{j+m-1}^{k-2l}\left(\frac{m_i+K_i}{\sqrt{\tau_i}}\right)$, $\tau_i=\left(m_i+K_i\right)^2-\left(K_i\Delta_i\right)^2$, and $u_i=2\lambda_i^2\bar\gamma_i$; $\Upsilon\left(\cdot,\cdot\right)$ is the lower incomplete gamma function \cite[eq. (8.350.1)]{b11}, $P\left(\cdot\right)$ is the associated Legendre function of the first kind \cite[eq. (8.702)]{b11}, and $\Gamma\left(\cdot\right)$ is the complete gamma function \cite[eq. (8.31)]{b11}. Besides, $K_{i}$, $\Delta_{i}$, $m_i$, and $\lambda^2_{i}=\frac{1}{2\left(1+K_i\right)}$ denote the fading parameters \cite{b8}.

On the other hand, for the Gaussian channel, i.e., $y_i=\sqrt{\gamma_i}s+n_i$, where $n_i$ is the normalized noise, one can express the mutual information of $M$-QAM signaling as \cite{b12,b22}
\begin{align}
&{{I}}_M\left(\gamma_i\right)=\log_2M-\frac{1}{M\pi}\sum\nolimits_{j=1}^{M}\int_{y\in{\mathbb C}}{{\exp}\left({-\left|{{y}}-\sqrt{\gamma_i}{{s_j}}\right|^2}\right)}\nonumber\\
&\times\log_2\left(\sum_{k=1}^{M}{{\exp}\left({\left|{{y}}-\sqrt{\gamma_i}{{s_j}}\right|^2-{{\left|{{y}}-\sqrt{\gamma_i}{{s_k}}\right|^2}}}\right)}\right){\rm{d}}y,\label{EQ2}
\end{align}
where $s_j$, $j\in\left\{1,2,\cdots,M\right\}$ denote the signals of $M$-QAM. Thus, the instantaneous secrecy rate can be written as \cite{b13}
\vspace{-5pt}
\begin{equation}
\label{EQ4}
{\mathcal I}_{\rm{s}}=I_{\rm{s}}\left(\gamma_{\rm{B}},\gamma_{\rm{E}}\right)=\max\left\{I_M\left(\gamma_{\rm{B}}\right)-I_M\left(\gamma_{\rm{E}}\right),0\right\}.
\vspace{-5pt}
\end{equation}
In the sequel, we will first discuss the properties of FDA systems and then derive the ASR and SOP to characterize the system performance.

\section{Discussion About The Beampattern}
\label{section3}
Using \eqref{equation1}, the beampattern can be written as \cite{b4}
\begin{equation}
\left|{\Theta}_{i}\right|=\frac{1}{N}{{\left| \frac{\sin N\pi \left( \frac{1}{2}\left[\sin {{\theta }_{i}}-\sin {{\theta }_{\rm B}}\right]-\frac{1}{c}{\Delta f}{\Delta r} \right)}{\sin \pi \left( \frac{1}{2}\left[\sin {{\theta }_{i}}-\sin {{\theta }_{\rm B}}\right]-\frac{1}{c}{\Delta f}{\Delta r} \right)} \right|}},
\end{equation}
where $\Delta r=r_{i}-r_{\rm B}$.
For a PA system, we have $\Delta f=0$ \cite{b3}, and thus $\left|{\Theta}_{\rm{E}}\right|=\left|{\Theta}_{\rm{B}}\right|=1$ always holds for ${\theta }_{\rm E}={\theta }_{\rm B}$, which means that a PA cannot improve the transmission security if Bob and Eve are in the same direction. By contrast, in an FDA system, we have $\Delta f>0$, and thus $\left|{\Theta}_{\rm{E}}\right|\leq\left|{\Theta}_{\rm{B}}\right|=1$ holds for ${\theta }_{\rm E}={\theta }_{\rm B}$. By properly setting the values of $\Delta f$ and $N$, an FDA can make $\left|{\Theta}_{\rm{E}}\right|<\left|{\Theta}_{\rm{B}}\right|=1$ and even $\left|{\Theta}_{\rm{E}}\right|=0$, which can reduce the SNR of Eve and thus enhance the transmission security. Therefore, FDA systems outperform PA systems in the range dimension (${\theta }_{\rm E}={\theta }_{\rm B}$). On the other hand, if ${\theta }_{\rm E}\neq{\theta }_{\rm B}$, $\left|{\Theta}_{\rm{E}}\right|$ can be designed to be smaller than $\left|{\Theta}_{\rm{B}}\right|$ regardless of $\Delta f$, and thus both PA and FDA systems can improve transmission security in the direction dimension.

For the sake of brevity, we set ${\theta }_{\rm E}={\theta }_{\rm B}$ to discuss more properties of the FDA. Under this set-up, we have $\left|{\Theta}_{i}\right|=\frac{1}{N}{{\left| \frac{\sin N\pi  \frac{{\Delta f}{\Delta r}}{c} }{\sin \pi  \frac{{\Delta f}{\Delta r}}{c} } \right|}}$ and find that the null points for $\left|{\Theta}_{\rm E}\right|=0$ along the range dimension satisfy $\Delta r=\frac{kc}{N\Delta f}$ ($k\in {\mathbb Z},~k\neq{mN},~m\in{\mathbb Z}$, $\mathbb Z$ denotes the integer set). Hence, the distance between two neighbour null points is $d_{\rm {null}}=\frac{c}{N\Delta f}$. To improve the secrecy performance, one effective method is to increase the number of null points of Eve or to reduce $d_{\rm {null}}$, which can be done by increasing $N$ or/and $\Delta f$. On the other hand, ${\mathcal L}\left(x\right)=\frac{1}{N}{{\left| \frac{\sin N\pi  \frac{{\Delta f}{x}}{c} }{\sin \pi  \frac{{\Delta f}{x}}{c} } \right|}}$ is a periodic function. In its first least positive period $\left(-\frac{c}{2\Delta f},\frac{c}{2\Delta f}\right)$, for $x$ that falls outsides the main-lobe $\left(-\frac{c}{N\Delta f},\frac{c}{N\Delta f}\right)$, we have ${\mathcal L}\left(x\right)<\frac{1}{N}\frac{1}{\sin\frac{\pi}{N}}$. We find that $\frac{1}{N}\frac{1}{\sin\frac{\pi}{N}}$ decreases monotonically with $N$, and thus using a large value of $N$ can effectively suppress the side-lobes of ${\mathcal L}\left(x\right)$. Thus, the received SNR of Eve that is located in the side-lobes can get decreased by increasing $N$, which can improve the security. In summary, the secrecy performance of FDA systems can get improved by increasing $N$ or/and $\Delta f$.

\section{Average Secrecy Rate}
\label{sec3}
\subsection{Explicit Analysis}
According to \eqref{EQ4}, the average secrecy rate is given by
\begin{align}
{\bar{{\mathcal I}}}_{s}
=\int_{0}^{+\infty}\int_{\gamma_{\rm{E}}}^{+\infty}\left[I_M\left(\gamma_{\rm{B}}\right)-I_M\left(\gamma_{\rm{E}}\right)\right]\nonumber\\
\times f_{\rm{B}}\left(\gamma_{\rm{B}}\right)f_{\rm{E}}\left(\gamma_{\rm{E}}\right){\rm{d}}\gamma_{\rm{B}}{\rm{d}}\gamma_{\rm{E}}.\label{EQ5}
\end{align}
{Yet, it is challenging to derive a closed-form expression of ${\bar{{\mathcal I}}}_{s}$ as the Gaussian integration in \eqref{EQ2} lacks any close-form solutions \cite{b12}. Fortunately, by exploiting \cite{b12}, $I_M\left(\gamma\right)$ can be written as
\begin{equation}
{{I}_M}\left(\gamma\right)=\log_2M\times\left(1-\sum\nolimits_{{j}=1}^{k_M}\zeta_j^{(M)}{\rm e}^{{-\vartheta_j^{(M)}\gamma}}\right)+E_M\left(\gamma\right),\nonumber
\end{equation}
where $k_M$, $\zeta_j^{(M)}$, and $\vartheta_j^{(M)}$ can be found in \cite[Table \uppercase\expandafter{\romannumeral1}]{b12}; $\zeta_j^{(M)}>0$, $\vartheta_j^{(M)}>0$, and $\sum_{{j}=1}^{k_M}\zeta_j^{(M)}=1$; $\lim_{\gamma\rightarrow0}E_M\left(\gamma\right)=\lim_{\gamma\rightarrow+\infty}E_M\left(\gamma\right)=0$. Define ${\hat{I}_M}\left(\gamma\right)\triangleq\log_2M\times\left(1-\sum_{{j}=1}^{k_M}\zeta_j^{(M)}{\rm e}^{{-\vartheta_j^{(M)}\gamma}}\right)$. Then, when $\gamma_{\rm B}>\gamma_{\rm E}$, the secrecy rate ${\mathcal I}_{s}=I_M\left(\gamma_{\rm B}\right)-I_M\left(\gamma_{\rm E}\right)$ satisfies $\left|\left[{\hat{I}_M}\left(\gamma_{\rm B}\right)-{{\hat I}_M}\left(\gamma_{\rm E}\right)\right]-{\mathcal I}_{s}\right|<{2}\mathfrak{T}^{(M)}$, where $\mathfrak{T}^{(M)}\triangleq\max_{0<\gamma<\infty}{\left|E_M\left(\gamma\right)\right|}$ can be found in \cite[Table \uppercase\expandafter{\romannumeral2}]{b12} and thus the ASR satisfies
\vspace{-3pt}
\begin{equation}
\left|{\bar{{\mathcal I}}}_{s}-\log_2M\times\sum_{{j}=1}^{k_M}\zeta_j^{(M)}{\mathcal{P}}\left(\vartheta_j^{(M)}\right)\right|<2\mathfrak{T}^{(M)}\Pr\left(\gamma_{\rm B}>\gamma_{\rm E}\right),\nonumber
\vspace{-3pt}
\end{equation}
where ${\mathcal{P}}\left(\vartheta\right)=\int_{0}^{+\infty}\int_{\gamma_{\rm{E}}}^{+\infty}\left({\rm e}^{{-\vartheta \gamma_{\rm{E}}}}-{\rm e}^{{-\vartheta \gamma_{\rm{B}}}}\right)f_{\rm{B}}\left(\gamma_{\rm{B}}\right)f_{\rm{E}}\left(\gamma_{\rm{E}}\right)\times{\rm{d}}\gamma_{\rm{B}}{\rm{d}}\gamma_{\rm{E}}$.
Based on \cite{b12}, we have $\mathfrak{T}^{(M)}={\mathcal O}\left(10^{-4}\right)$, which suggests that ${\hat{I}_M}\left(\gamma_{\rm B}\right)-{{\hat I}_M}\left(\gamma_{\rm E}\right)$ can provide a good approximation of ${{{\mathcal I}}}_{s}$.
Moreover, due to the facts of $\mathfrak{T}^{(M)}={\mathcal O}\left(10^{-4}\right)$ and $\Pr\left(\gamma_{\rm B}>\gamma_{\rm E}\right)<1$, it makes sense to approximate ${\bar{{\mathcal I}}}_{s}$ as ${\hat{{\mathcal I}}}_{s}=\log_2M\times\sum_{{j}=1}^{k_M}\zeta_j^{(M)}{\mathcal{P}}\left(\vartheta_j^{(M)}\right)$, and ${\hat{{\mathcal I}}}_{s}-2\mathfrak{T}^{(M)}\Pr\left(\gamma_{\rm B}>\gamma_{\rm E}\right)$ serves as a lower bound of ${\bar{{\mathcal I}}}_{s}$.}

\setcounter{TempEqCnt}{\value{equation}}
\setcounter{equation}{8}
\begin{figure*}[hb]
\vspace{-15pt}
\hrulefill
\vspace{-5pt}
\begin{align}
{\hat{{\mathcal I}}}_{{s}}=&\sum_{{q}=1}^{k_M}\sum_{k=0}^{\infty}\frac{\log_2M\times\zeta_q^{(M)}m_{\rm E}^{m_{\rm E}}K_{\rm E}^kd_k^{\left(\rm E\right)}}{\Gamma\left(m_{\rm E}\right)k!\left(1+u_{\rm E}\vartheta_q^{(M)}\right)^{k+1}}-\log_2M\times\sum_{{q}=1}^{k_M}\sum_{j=0}^{\infty}\sum_{k=0}^{\infty}\frac{\zeta_q^{(M)}m_{\rm B}^{m_{\rm B}}m_{\rm E}^{m_{\rm E}}K_{\rm B}^jd_j^{\left({\rm B}\right)}K_{\rm E}^kd_k^{\left({\rm E}\right)}}{\Gamma\left(m_{\rm B}\right)\Gamma\left(m_{\rm E}\right)j!k!}\Biggg[\frac{1}{\left(u_{\rm B}\vartheta_q^{(M)}+1\right)^{j+1}}\nonumber\\
&-\sum_{l=0}^{k}\frac{\Gamma\left(l+j+1\right)u_{\rm B}^{-1-j}u_{\rm E}^{-l}}{l!j!\left(\vartheta_q^{(M)}+\frac{1}{u_{\rm B}}+\frac{1}{u_{\rm E}}\right)^{l+j+1}}+\frac{1}{\left(u_{\rm E}\vartheta_q^{(M)}+1\right)^{k+1}}
-\sum_{l=0}^{j}\frac{\Gamma\left(l+k+1\right)u_{\rm E}^{-1-k}u_{\rm B}^{-l}}{l!k!\left(\vartheta_q^{(M)}+\frac{1}{u_{\rm B}}+\frac{1}{u_{\rm E}}\right)^{l+k+1}}
\Biggg]\label{eq19}
\end{align}
\vspace{-10pt}
\end{figure*}
\setcounter{equation}{\value{TempEqCnt}}
\setcounter{equation}{7}

To facilitate the derivation, we re-express ${\mathcal{P}}\left(\vartheta\right)$ as
${\mathcal{P}}\left(\vartheta\right)=\int_{0}^{+\infty}{\mathcal{A}}_3\left(x\right){\rm d}x-\int_{0}^{+\infty}{\mathcal{A}}_1\left(x\right){\rm d}x-\int_{0}^{+\infty}{\mathcal{A}}_2\left(x\right){\rm d}x$, where ${\mathcal{A}}_1\left(x\right)=\exp\left(-\vartheta x\right)f_{\rm{B}}\left(x\right)F_{\rm{E}}\left(x\right){\rm{d}}x$,
${\mathcal{A}}_2\left(x\right)=\exp\left(-\vartheta x\right)f_{\rm{E}}\left(x\right)F_{\rm{B}}\left(x\right)$, and ${\mathcal{A}}_3\left(x\right)=\exp\left(-\vartheta x\right)f_{\rm{E}}\left(x\right)$. Since $j+1$ are positive integers, $F_i\left(\gamma_i\right)$ can be written as \cite[eq. (8.352.6)]{b11}
\vspace{-5pt}
\begin{equation}
\label{eq9}
F_i\left(\gamma_i\right)=\frac{m_i^{m_i}}{\Gamma\left(m_i\right)}\sum_{j=0}^{\infty}\frac{K_i^jd^{\left(i\right)}_j}{j!}
\left(1-\sum_{q=0}^{j}\frac{{\rm e}^{-{\gamma_i}/{u_i}}}{\gamma_i^{-q}u_i^qq!}\right),
\vspace{-5pt}
\end{equation}
Then, we substitute \eqref{eq1a} and \eqref{eq9} into ${\mathcal{P}}\left(\vartheta\right)$, and apply the integral identity from \cite[eq. (3.326.2)]{b11} to solve the resultant integrals. Taken together, the expression of ${\hat{{\mathcal I}}}_{{s}}$ is summarized in \eqref{eq19} at the bottom of this page. Theoretically, both $j$ and $k$ in \eqref{eq19} take all the integers in the range $\left[0,+\infty\right]$, which is impractical for exact calculation. Fortunately, the first tens of terms of $j$ and $k$ contribute most of the values to the results of the computation and setting $j, k\in\left[0,40\right]$ can obtain extremely high calculation accuracy \cite{b8}. Hence, we apply this technique in calculating ASR and SOP.

\setcounter{equation}{9}

\subsection{Asymptotic Analysis}
\label{Section3B}
To show more system insights, this section will present an asymptotic analysis on the ASR when $\bar\gamma_{\rm B}$ tends to infinity. {Based on \eqref{EQ5}, we can rewrite ${\bar{{\mathcal I}}}_{s}$ as
\begin{align}
&{\bar{{\mathcal I}}}_{s}=\int_{0}^{\infty}I_{M}\left(\gamma\right)\left[f_{\rm B}\left(\gamma\right)F_{\rm E}\left(\gamma\right)+f_{\rm E}\left(\gamma\right)F_{\rm B}\left(\gamma\right)\right]{\rm d}\gamma\nonumber\\
&-\int_{0}^{\infty}I_{M}\left(\gamma\right)f_{\rm E}\left(\gamma\right){\rm d}\gamma=-\int_{0}^{\infty}I_{M}\left(\gamma\right)f_{\rm E}\left(\gamma\right){\rm d}\gamma\nonumber\\
&+\int_{0}^{\infty}I_{M}\left(\gamma\right){\rm d}\left(F_{\rm B}\left(\gamma\right)F_{\rm E}\left(\gamma\right)\right)=\log_2{M}-\int_{0}^{\infty}F_{\rm B}\left(\gamma\right)\nonumber\\
&\times F_{\rm E}\left(\gamma\right){\rm d}I_{M}\left(\gamma\right)-\int_{0}^{\infty}I_{M}\left(\gamma\right)f_{\rm E}\left(\gamma\right){\rm d}\gamma.
\end{align}
By \cite[eq. (8.354.1)]{b11}, $\lim\limits_{x\rightarrow0}\Upsilon\left(s,x\right)=\frac{x^s}{s}+{o}\left(x^s\right)$, where $o\left(\cdot\right)$ denotes the higher order term. Hence, for fixed $\gamma$, $F_{\rm B}\left(\gamma\right)=\frac{\chi_{\rm B}\gamma}{\bar\gamma_{\rm B}}+{o}\left(\frac{1}{\bar\gamma_{\rm B}}\right)$ as $\bar\gamma_{\rm B}\rightarrow\infty$, where $\chi_{\rm B}=\frac{m_{\rm B}^{m_{\rm B}}d_0^{\left(\rm B\right)}}{2\lambda_{\rm B}^2\Gamma\left(m_{\rm B}\right)}$. Define ${\mathcal S}_M\left(\gamma\right)\triangleq\frac{{\rm d}I_{M}\left(\gamma\right)}{{\rm d}\gamma}$. Then the asymptotic ASR is given by
\vspace{-3pt}
\begin{equation}
{\bar{{\mathcal I}}}_{s}^{\infty}={\bar{{\mathsf I}}}_{s}^{\infty}-\int_{0}^{\infty}\frac{\chi_{\rm B}}{\bar\gamma_{\rm B}}\gamma F_{\rm E}\left(\gamma\right){\mathcal S}_M\left(\gamma\right){{\rm d}\gamma}+{o}\left({\bar\gamma_{\rm B}^{-1}}\right),
\vspace{-3pt}
\end{equation}
where ${\bar{{\mathsf I}}}_{s}^{\infty}=\log_2{M}-\int_{0}^{\infty}I_{M}\left(\gamma\right)f_{\rm E}\left(\gamma\right){\rm d}\gamma$ is a constant that is solely determined by Eve's channel condition. By approximating ${I}_{M}\left(\gamma\right)$ as $\hat{I}_{M}\left(\gamma\right)$, we can approximate ${\bar{{\mathsf I}}}_{s}^{\infty}$ as ${\bar{{\mathsf I}}}_{s}^{\infty}\approx\sum_{{q}=1}^{k_M}\sum_{k=0}^{\infty}\frac{\log_2M\times m_{\rm E}^{m_{\rm E}}\zeta_j^{(M)}K_{\rm E}^kd_k^{\left(\rm E\right)}}{k!\Gamma\left(m_{\rm E}\right)\left(1+\vartheta_j^{(M)}u_{\rm E}\right)^{k+1}}$ aided with \cite[eq. (3.326.2)]{b11}. Besides, ${\bar{{\mathsf I}}}_{s}^{\infty}$ can be also calculated by methods of numerical integration. Moreover, we have $\lim_{\gamma\rightarrow0}{\mathcal S}_M\left(\gamma\right)=\frac{1}{\ln{2}}$ \cite{b22} and $\lim_{\gamma\rightarrow\infty}{\mathcal S}_M\left(\gamma\right)=o\left({\rm e}^{-d_M\gamma}\right)$ ($d_M=\frac{3}{2\left(M-1\right)}>0$) \cite{b26}, which together with \cite[Section 4.2]{b23}, yields $\int_{0}^{\infty}{\gamma}F_{\rm E}\left(\gamma\right){\mathcal S}_M\left(\gamma\right){{\rm d}\gamma}<\infty$. Thus, we obtain ${\bar{{\mathcal I}}}_{s}^{\infty}={\bar{{\mathsf I}}}_{s}^{\infty}-{\Psi_M}{\bar\gamma_{\rm B}^{-1}}+{o}\left({\bar\gamma_{\rm B}^{-1}}\right)$ or $\left|{\bar{{\mathcal I}}}_{s}^{\infty}-{\bar{{\mathsf I}}}_{s}^{\infty}\right|={\mathcal O}{\left({\bar\gamma_{\rm B}^{-1}}\right)}$, where $\Psi_M=\chi_{\rm B}\int_{0}^{\infty}\gamma F_{\rm E}\left(\gamma\right){\mathcal S}_M\left(\gamma\right){{\rm d}\gamma}$ is a constant that can be calculated numerically with the aid of \cite{b22}. We find that as $\bar\gamma_{\rm B}$ increases, the ASR converges to a constant that is solely determined by Eve's channel quality and the rate of convergence is ${\mathcal O}{\left({\bar\gamma_{\rm B}^{-1}}\right)}$. In contrast, for Gaussian inputs, the ASR scales with $\bar\gamma_{\rm B}$ at the rate of ${\mathcal O}{\left(\log\bar\gamma_{\rm B}\right)}$ \cite{b22} which suggests the performance gap between Gaussian and finite-alphabet inputs.}

\section{Secrecy Outage Probability}
\subsection{Explicit Analysis}
The SOP is defined as the probability when the instantaneous secrecy rate is lower than a preset constant value $R_{s}>0$ \cite{b13}. By its definition, we formulate the SOP as
\vspace{-5pt}
\begin{equation}
\label{EQ23}
P_{\rm{o}}\left(R_{s}\right)=\Pr\left(I_{\rm{s}}\left(\gamma_{\rm{B}},\gamma_{\rm{E}}\right)<R_{s}\right).
\vspace{-5pt}
\end{equation}
According to \eqref{EQ4}, \eqref{EQ23} can be further simplified to
\vspace{-5pt}
\begin{equation}
P_{\rm{o}}\left(R_{s}\right)=\Pr\left({\mathcal I}_{s}<R_{s},\gamma_{\rm B}>\gamma_{\rm E}\right)+\Pr\left(\gamma_{\rm B}<\gamma_{\rm E}\right).\label{EQ19}
\vspace{-5pt}
\end{equation}
Define $I_M^{-1}\left(\cdot\right)$ as the inverse function of $I_M\left(\cdot\right)$. {Although $I^{-1}_{M}\left(\cdot\right)$ lacks an explicit expression, its value can be found via a simple bisection search.} The following lemma captures the main result of SOP:
\begin{lemma}
\vspace{-5pt}
  \label{Lemma1}
  Define ${\mathcal H}\triangleq I_M^{-1}\left(\log_2 M-R_{s}\right)$ and $\phi_M\left(\gamma\right)\triangleq{I_M^{-1}\left(R_{s}+I_M\left(\gamma\right)\right)}$. Then the SOP with discrete inputs is
  \vspace{-5pt}
  \begin{equation}
  P_{\rm o}\left(R_{s}\right)=1-F_{\rm{E}}\left(\mathcal H\right)
  +\int_{0}^{\mathcal H}F_{\rm{B}}\left(\phi_M\left(\gamma\right)\right)f_{\rm{E}}\left(\gamma\right){\rm{d}}\gamma.\label{EQ21}
  \vspace{0pt}
  \end{equation}
  \end{lemma}
  \vspace{-5pt}
\begin{IEEEproof}
Please see the appendix.
\end{IEEEproof}

Inserting \eqref{eq1a} and \eqref{eq1b} into \eqref{EQ21} yields
\begin{equation}
\begin{split}
P_{\rm{o}}\left(R_{s}\right)=&1-\frac{m_{\rm E}^{m_{\rm E}}}{\Gamma\left(m_{\rm E}\right)}\sum_{j=0}^{\infty}\frac{K_{\rm E}^jd^{\left({\rm E}\right)}_j}{j!j!}\Upsilon\left(j+1,\frac{{\mathcal H}}{u_{\rm E}}\right)\\
&+\sum_{j=0}^{\infty}\sum_{k=0}^{\infty}\frac{m_{\rm B}^{m_{\rm B}}m_{\rm E}^{m_{\rm E}}K_{\rm B}^jd_j^{\left({\rm B}\right)}K_{\rm E}^kd_k^{\left({\rm E}\right)}}{\Gamma\left(m_{\rm B}\right)\Gamma\left(m_{\rm E}\right)j!j!k!k!u_{\rm E}^2}{\mathsf P}_k,\label{EQ31}
\end{split}
\end{equation}
where ${\mathsf P}_k=\int_{0}^{\mathcal H}{\Upsilon\left(j+1,\frac{{\phi_M\left(\gamma\right)}}{u_{\rm B}}\right)}{\gamma^{k}\exp\left(-\frac{1}{u_{\rm E}}\gamma\right)}{\rm{d}}\gamma$. Furthermore, the integral in ${\mathsf P}_k$ can be effectively evaluated by the Gauss-Legendre quadrature rule \cite{b14}, namely
\vspace{-5pt}
\begin{equation}
{\mathsf P}_k\approx\frac{{\mathcal H}}{2}\sum_{i=1}^{V}t_i\frac{\Upsilon\left(j+1,\frac{1}{u_{\rm B}}{\phi_M\left(\frac{\mathcal H}{2}\omega_i+\frac{\mathcal H}{2}\right)}\right)}{\left(\frac{\mathcal H}{2}\omega_i+\frac{\mathcal H}{2}\right)^{-k}\exp\left(\frac{1}{u_{\rm E}}\left(\frac{\mathcal H}{2}\omega_i+\frac{\mathcal H}{2}\right)\right)},
\vspace{-5pt}
\end{equation}
where $t_i$ and $\omega_i$ can be found in \cite[Table 25.4]{b14}. Note that a larger value of $V$ indicates higher approximation precision and $V$ in this paper is set to 30 to achieve a promising accuracy.

\subsection{Asymptotic Analysis}
{As stated before, $\lim_{{\bar\gamma_{\rm B}\rightarrow\infty}}F_{\rm B}\left(\gamma\right)=\chi_{\rm B}\frac{\gamma}{\bar\gamma_{\rm B}}+{o}\left(\frac{1}{\bar\gamma_{\rm B}}\right)$ holds for fixed $\gamma$. Thus, as ${\bar\gamma_{\rm B}\rightarrow\infty}$, \eqref{EQ21} can be written as $P_{\rm{o}}^{\infty}\left(R_{s}\right)
=1-F_{\rm{E}}\left(\mathcal H\right)+{\Phi_M}{\bar\gamma_{\rm B}^{-1}}+{o}\left({\bar\gamma_{\rm B}^{-1}}\right)$, where $\Phi_M=\int_{0}^{\mathcal H}\chi_{\rm B}\phi_M\left(\gamma\right)f_{\rm{E}}\left(\gamma\right){\rm{d}}\gamma$. By applying the change of the variable $\gamma\rightarrow \rho_{M}\left(x\right)=I_{M}^{-1}\left(I_M\left(x\right)-R_s\right)$, we obtain $\Phi_M=\chi_{\rm B}\int_{I_{M}^{-1}\left(R_s\right)}^{+\infty}\frac{xf_{\rm E}\left(\rho_{M}\left(x\right)\right)}{S_M\left(\rho_{M}\left(x\right)\right)}S_M\left(x\right){\rm d}x$. Moreover, we have $\lim_{x\rightarrow I_{M}^{-1}\left(R_s\right)}\frac{f_{\rm E}\left(\rho_{M}\left(x\right)\right)}{S_M\left(\rho_{M}\left(x\right)\right)}{\mathcal S}_M\left(x\right)=0$ \cite{b22} and $\lim_{x\rightarrow\infty}\frac{f_{\rm E}\left(\rho_{M}\left(x\right)\right)}{S_M\left(\rho_{M}\left(x\right)\right)}{\mathcal S}_M\left(x\right)=o\left({\rm e}^{-d_Mx}\right)$ ($d_M>0$) \cite{b26}, which together with \cite[Section 4.2]{b23}, yields $\Phi_M<\infty$. Note that $\Phi_M$ can be calculated by the Gauss-Legendre quadrature rule. We find that as $\bar\gamma_{\rm B}$ increases, the SOP converges to $1-F_{\rm{E}}\left(\mathcal H\right)>0$. Besides, the rate of convergence is ${\mathcal O}{\left({\bar\gamma_{\rm B}^{-1}}\right)}$ or in other words, the order of convergence equals one. Based on \cite{b24}, for Gaussian inputs, $P_{\rm{o}}\left(R_{s}\right)$ converges to zero as $\bar\gamma_{\rm B}$ increases which indicates the performance gap between Gaussian and finite-alphabet inputs. Yet, by \cite{b24}, the order of convergence (or the secrecy diversity order) for Gaussian inputs equals one which is the same as that for discrete inputs.}

\section{Simulation}
\label{sec4}
\begin{figure}[!t]
    \centering
    \subfigbottomskip=0pt
	\subfigcapskip=-5pt
\setlength{\abovecaptionskip}{0pt}
    \subfigure[ASR versus $r_{\rm B}$.]
    {
        \includegraphics[height=0.13\textwidth]{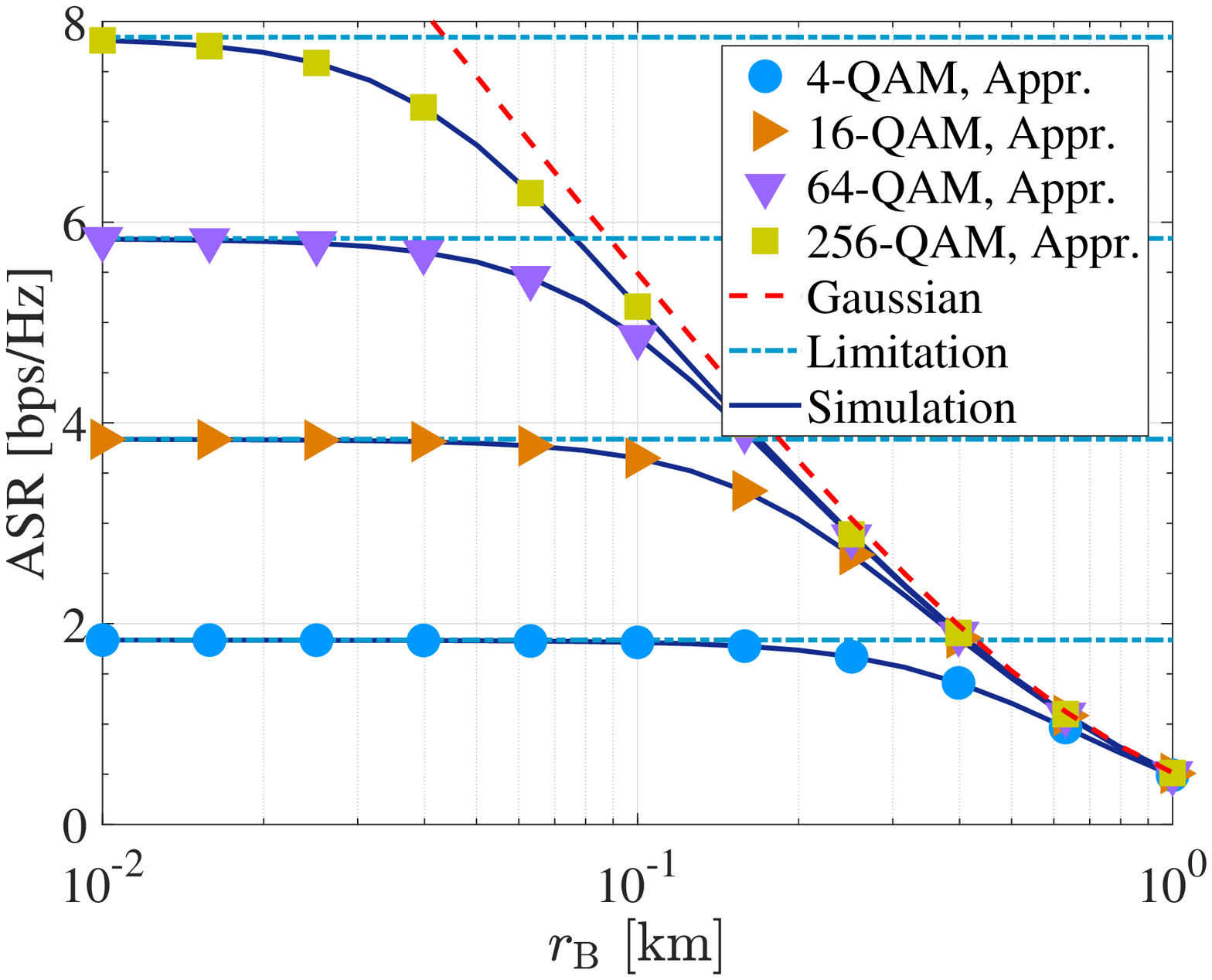}
	   \label{fig1a}	
    }
   \subfigure[$\Delta{\bar{{\mathcal I}}}_{s}^{\infty}$ versus $r_{\rm B}$.]
    {
        \includegraphics[height=0.13\textwidth]{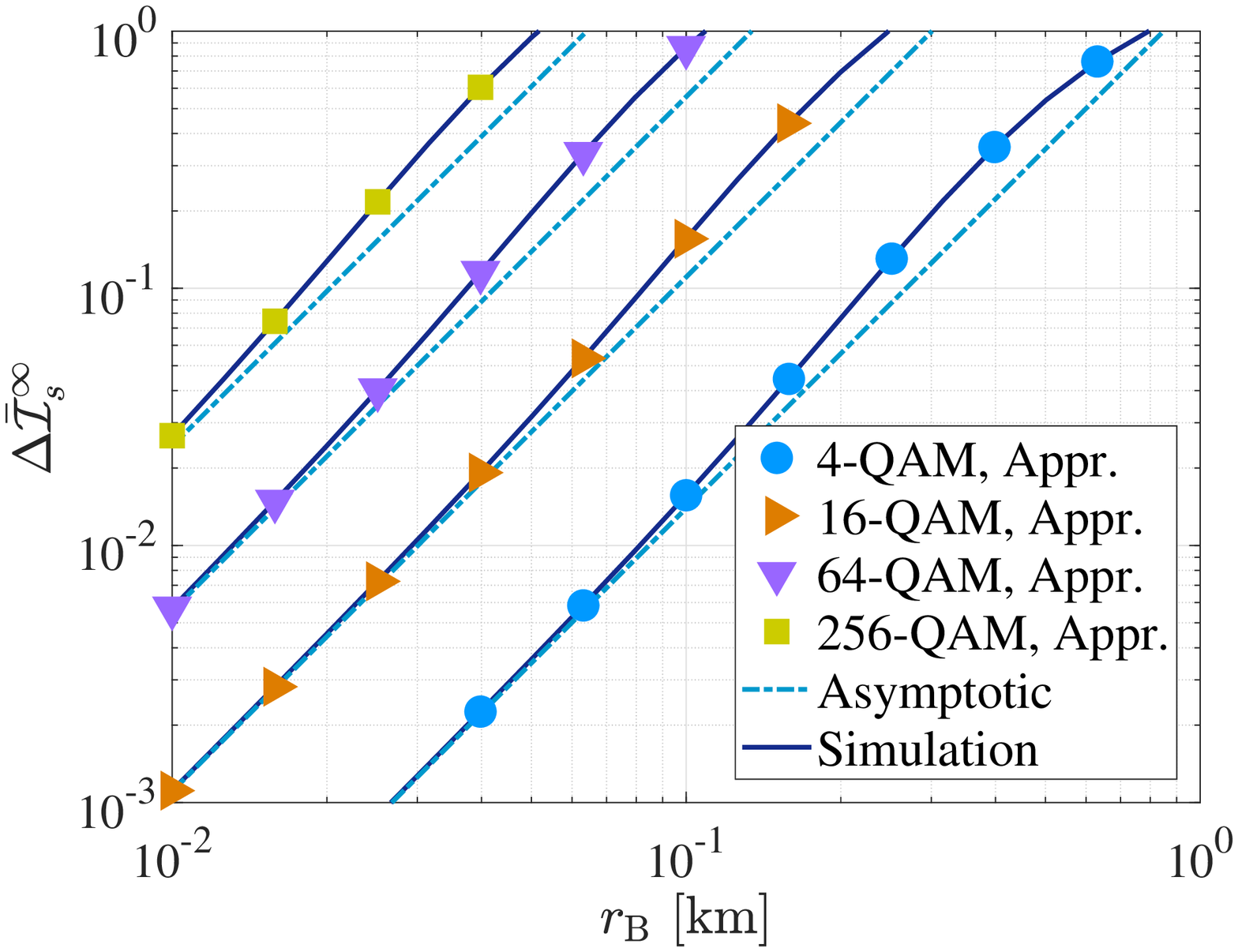}
	   \label{fig1b}	
    }
\caption{ASR of $M$-QAM versus $r_{\rm B}$ for $\left(m_{\rm B},\Delta_{\rm B},K_{\rm B}\right)=\left(2,0.4,10\right)$, $\left(m_{\rm E},\Delta_{\rm E},K_{\rm E}\right)=\left(5,0.35,5\right)$, $P=10$ dBW, $N=50$, $\Delta f=1$ kHz, $\theta_{\rm B}=\theta_{\rm E}={\ang{20}}$, and $r_{\rm E}=1.5$ km.}
    \label{figure1}
    \vspace{-15pt}
\end{figure}

\begin{figure}[!t]
    \centering
    \subfigbottomskip=0pt
	\subfigcapskip=-5pt
\setlength{\abovecaptionskip}{0pt}
    \subfigure[$R_{s}=1.8$ bps/Hz.]
    {
        \includegraphics[width=0.17\textwidth]{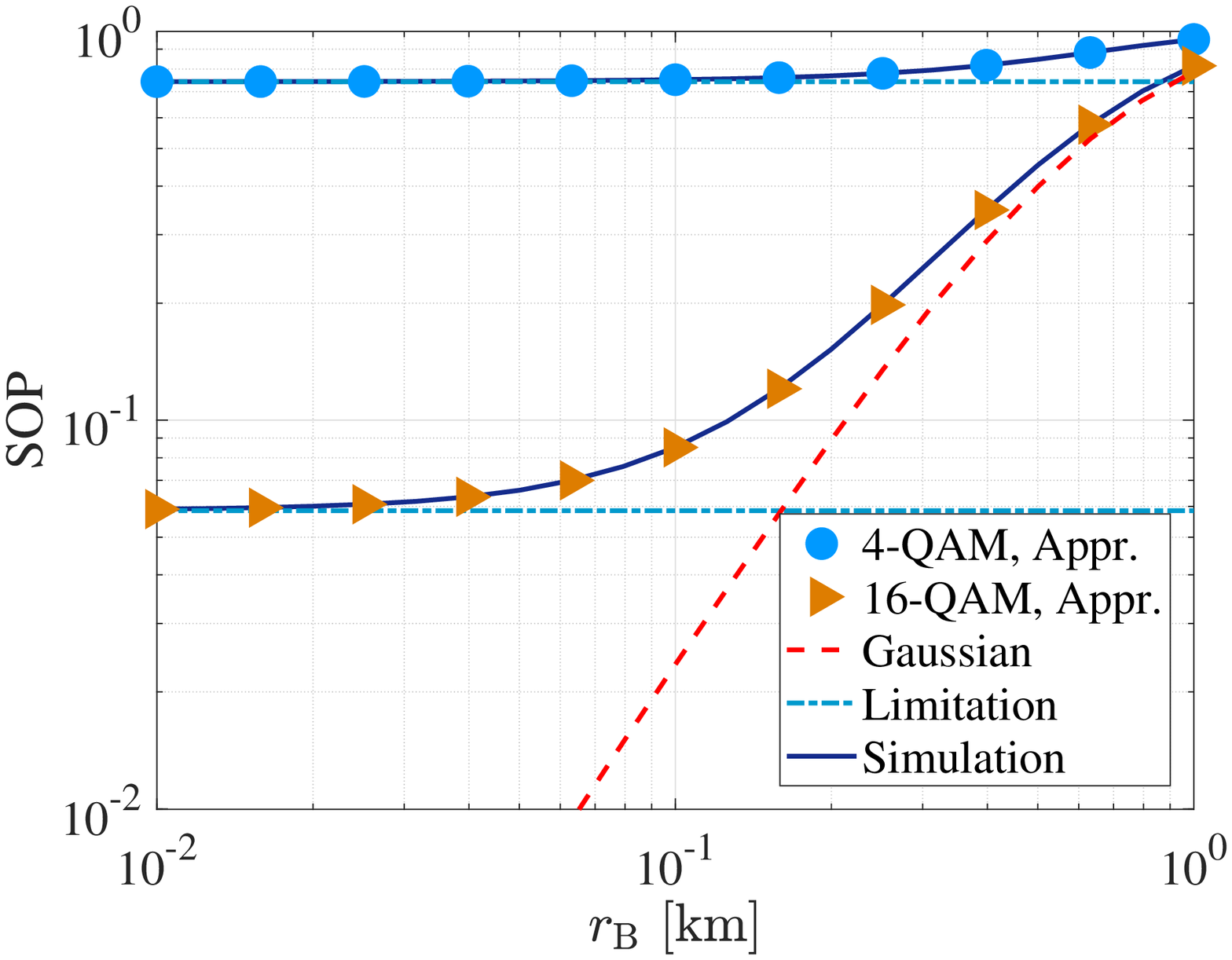}
	   \label{fig2a}	
    }
   \subfigure[$R_{s}=4.8$ bps/Hz.]
    {
        \includegraphics[width=0.17\textwidth]{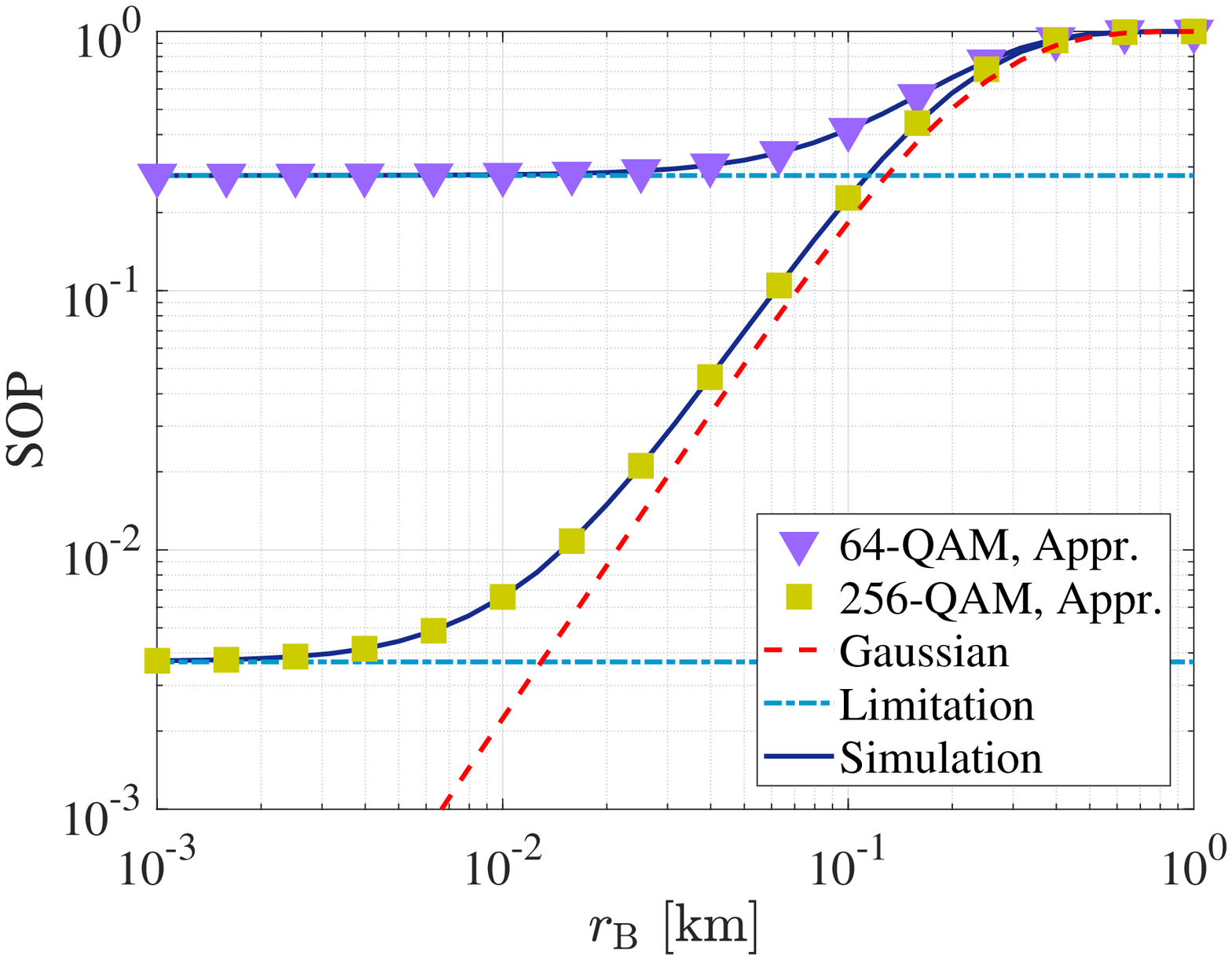}
	   \label{fig2b}	
    }
\caption{SOP of $M$-QAM versus $r_{\rm B}$ for $\left(m_{\rm B},\Delta_{\rm B},K_{\rm B}\right)=\left(1,1,1\right)$, $\left(m_{\rm E},\Delta_{\rm E},K_{\rm E}\right)=\left(1,1,20\right)$, $P=17$ dBW, $N=50$, $\Delta f=1$ kHz, $\theta_{\rm B}=\theta_{\rm E}={\ang{20}}$, and $r_{\rm E}=1.5$ km.}
    \label{figure2}
    \vspace{-20pt}
\end{figure}
To verify the derived analytical results, this part provides some numerical simulations. In the simulation, $f_{0}=28$ GHz \cite{b8} and $\sigma_{\rm B}^2=\sigma_{\rm E}^2=-140$ dBm. Furthermore, the free-space model is $10\log_{10} A_i\left(f_0,r_i\right)=32.5+20\log_{10}\left[f_0\left({\rm{MHz}}\right)\right]+20\log_{10}\left[r_i\left({\rm {km}}\right)\right]$ \cite[eq. (34)]{b2}.

{\figurename} \ref{fig1a} compares the simulated and approximated ASR of $M$-QAM versus $r_{\rm B}$. As shown, the simulations match perfectly with the approximations. Note that the free-space path loss is considered in this work, and thus the average received SNR of Bob is smaller for a larger distance $r_{\rm B}$. Therefore, as $r_{\rm B}$ decreases, the ASR gradually converges to its limitation, namely $\bar{\mathsf I}^{\infty}_{s}$, which consists with the results shown in {\figurename} \ref{fig1a}. For comparison, the ASR of Gaussian signal is also plotted. {By \cite{b15}, Gaussian signals can maximize the difference of mutual information over the main and eavesdropper's channels. Thus, Gaussian inputs outperform the discrete inputs in terms of ASR.} Moreover, as shown in {\figurename} \ref{fig1a}, a lower modulation order can achieve nearly the same secrecy level as a higher modulation order in low SNR regimes. The reason lies in that {$I_M\left(\gamma\right)$ has similar asymptotic behavior regardless of $M$} when $\gamma\rightarrow0$ \cite[eq. (92)]{b22}, and thus different modulation schemes can achieve virtual the same secrecy performance in low SNR regimes. Then we plot $\Delta{\bar{{\mathcal I}}}_{s}^{\infty}={\bar{{\mathsf I}}}_{s}^{\infty}-{\bar{{\mathcal I}}}_{s}^{\infty}$ versus $r_{\rm B}$ in {\figurename} \ref{fig1b} to verify the asymptotic analysis on ${\bar{{\mathcal I}}}_{s}$. In fact, we have $\Delta{\bar{{\mathcal I}}}_{s}^{\infty}={\Psi_M}{\bar\gamma_{\rm B}^{-1}}+{o}\left({\bar\gamma_{\rm B}^{-1}}\right)$. As shown, the derived asymptotic results match the simulated results perfectly in high SNR regimes. {\figurename} \ref{figure2} verifies the accuracy of derived expressions of the SOP. We observe that as $r_{\rm B}$ decreases, the SOP gradually converges to its limitation, namely $1-F_{\rm E}\left(\mathcal H\right)$. Besides, as shown in {\figurename} \ref{figure2}, a higher modulation order yields a smaller value of $1-F_{\rm E}\left(\mathcal H\right)$. We comment that the asymptotic behaviour of $P_{\rm{o}}\left(R_{s}\right)$ is similar as that of ${\bar{{\mathcal I}}}_{s}$. Due to the page limitation, we do not present the simulations of $P_{\rm{o}}^{\infty}\left(R_{s}\right)$.

\begin{figure}[!t]
    \centering
    \subfigbottomskip=1pt
	\subfigcapskip=-5pt
\setlength{\abovecaptionskip}{0pt}
    \subfigure[$r_{\rm{E}}=2.5$ km, $N=150$.]
    {
        \includegraphics[width=0.17\textwidth]{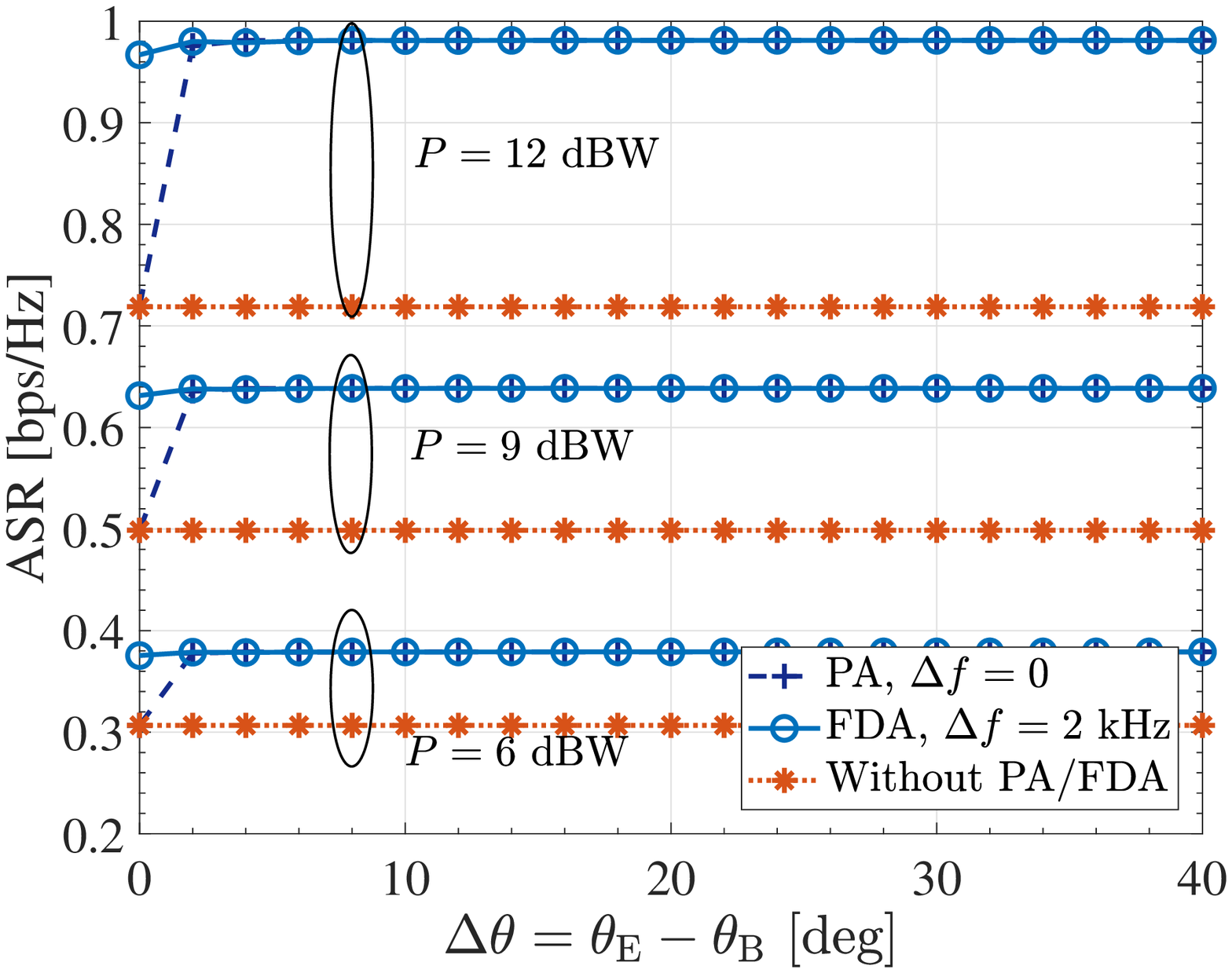}
	   \label{fig3a}	
    }\hspace{5pt}
   \subfigure[$\theta_{\rm E}=\ang{20}$, $N=100$.]
    {
        \includegraphics[width=0.17\textwidth]{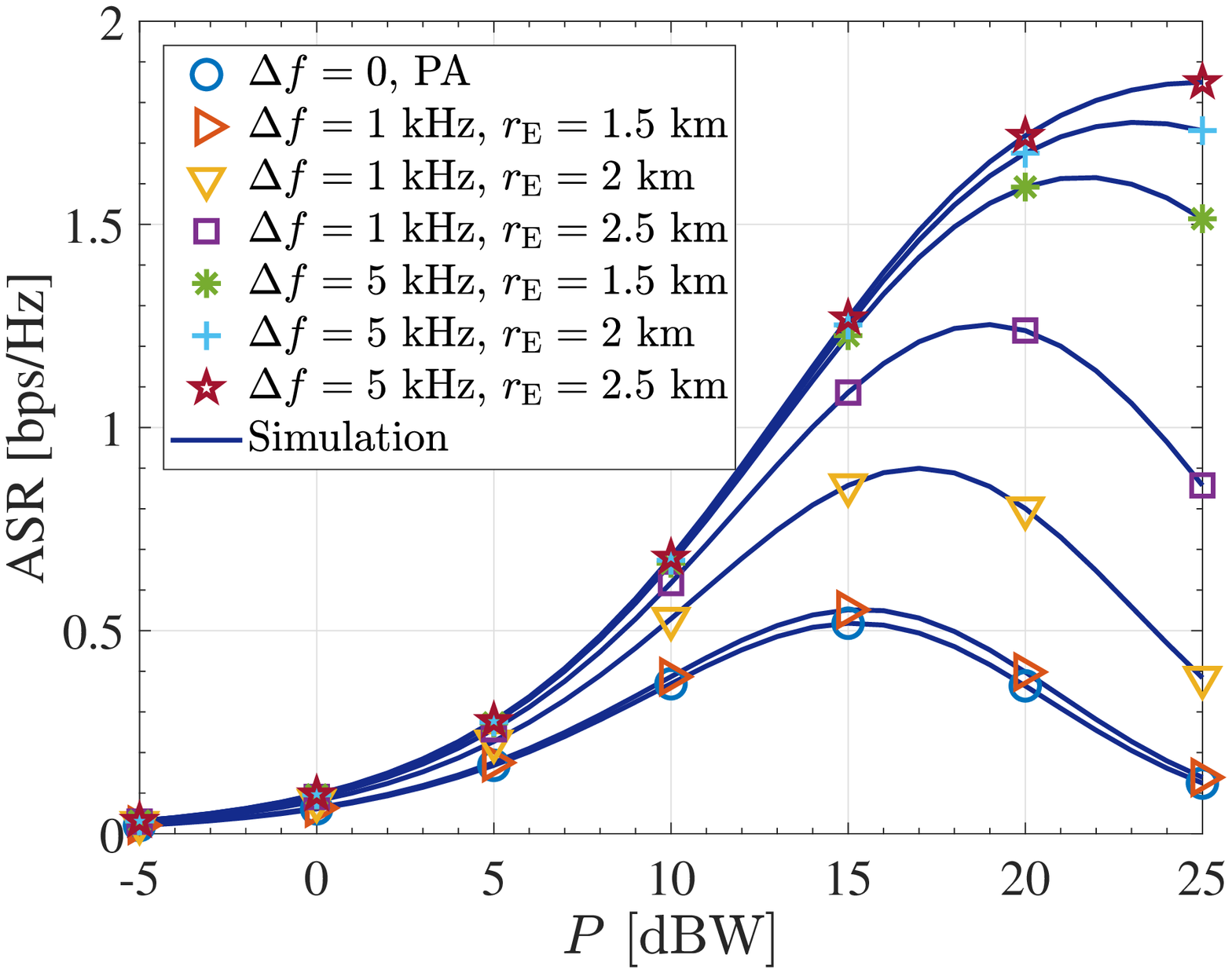}
	   \label{fig3b}	
    }
\caption{ASR of $4$-QAM for $\theta_{\rm B}=\ang{20}$, $r_{\rm B}=1$ km, $\left(m_{\rm B},\Delta_{\rm B},K_{\rm B}\right)=\left(2,0.4,10\right)$, and $\left(m_{\rm E},\Delta_{\rm E},K_{\rm E}\right)=\left(5,0.35,5\right)$.}
    \label{figure3}
    \vspace{-17pt}
\end{figure}

\begin{figure}[!t]
    \centering
    \subfigbottomskip=1pt
	\subfigcapskip=-5pt
\setlength{\abovecaptionskip}{0pt}
    \subfigure[$m_{\rm{B}}=m_{\rm{E}}=5$.]
    {
        \includegraphics[width=0.17\textwidth]{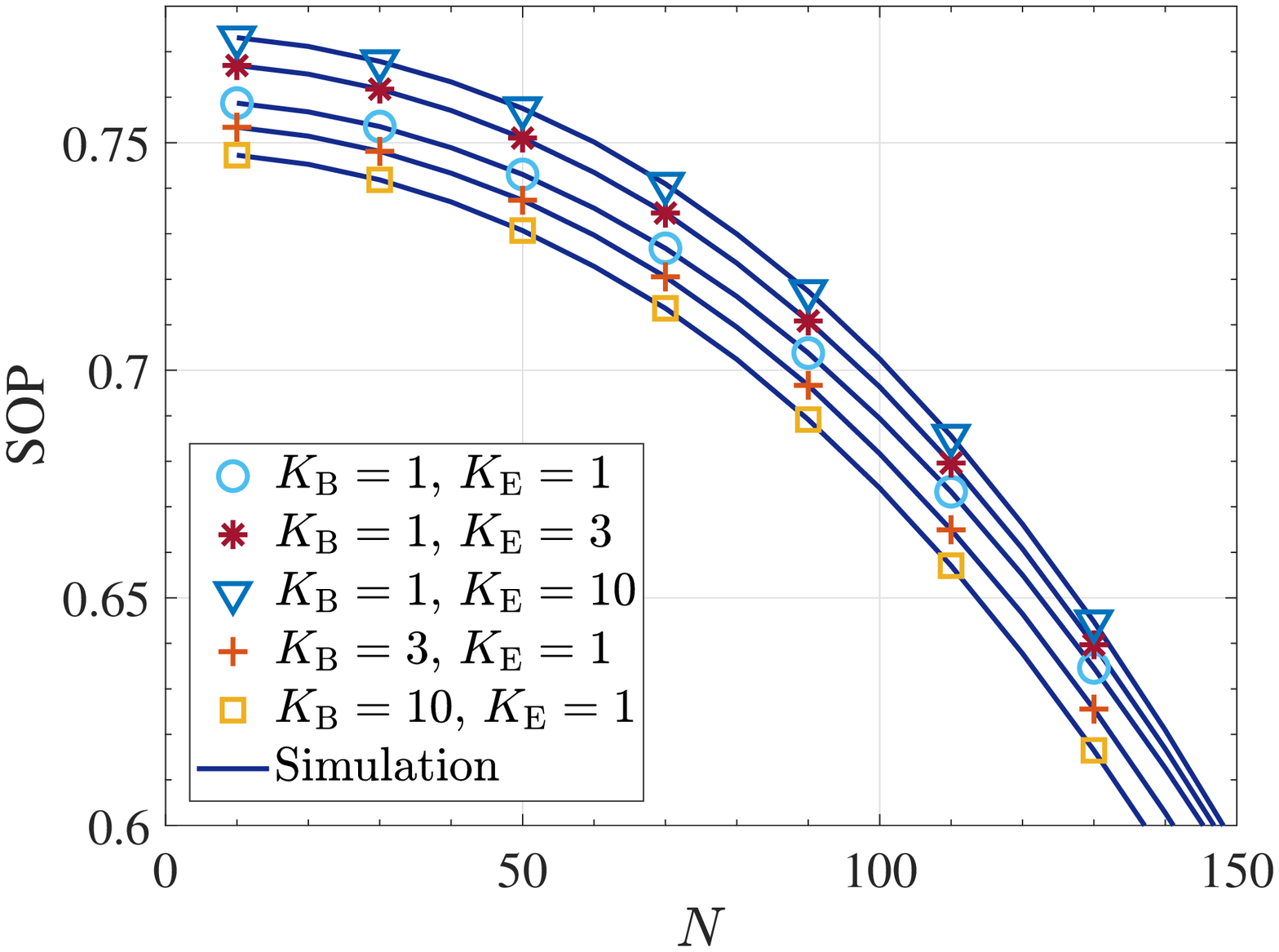}
	   \label{fig4a}	
    }\hspace{5pt}
   \subfigure[$K_{\rm{B}}=K_{\rm{E}}=3$.]
    {
        \includegraphics[width=0.17\textwidth]{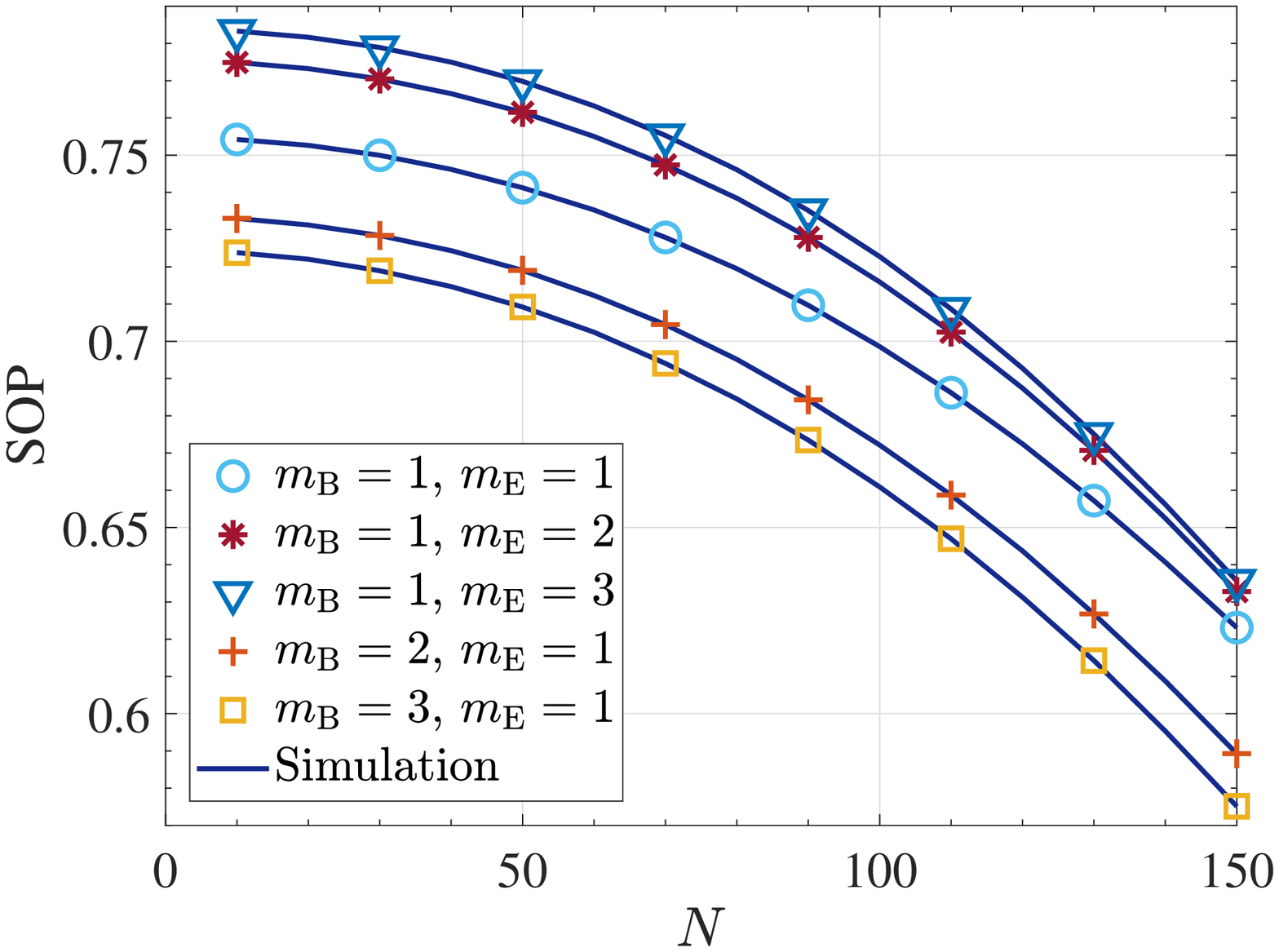}
	   \label{fig4b}	
    }
\caption{SOP of $4$-QAM for $R_{s}=1$ bps/Hz, $\Delta_{\rm B}=\Delta_{\rm E}=0.9$, $P=15$ dBW, $\Delta f=2$ kHz, $\theta_{\rm B}=\theta_{\rm E}={\ang{20}}$, $r_{\rm B}=1$ km, and $r_{\rm E}=1.5$ km.}
    \label{figure4}
    \vspace{-17pt}
\end{figure}

{\figurename} \ref{fig3a} plots the ASR achieved by PA and FDA systems versus $\Delta \theta=\theta_{\rm E}-\theta_{\rm B}$. For comparison, the ASR achieved in a FTR fading channel without using antenna array is also presented. As shown, both PA and FDA systems can improve the transmission security in the direction dimension ($\theta_{\rm E}\neq\theta_{\rm B}$). Yet, in the range dimension ($\theta_{\rm E}=\theta_{\rm B}$), only the FDA can improve the secrecy performance, which means FDA systems can achieve two-dimensional secure transmission. {\figurename} \ref{fig3b} plots the ASR achieved by FDA systems when $\theta_{\rm E}=\theta_{\rm B}$ for selected values of $\Delta f$. As shown, the FDA outperforms the PA in the range dimension and a larger value of $\Delta f$ yields a better secrecy performance, which verifies the discussion in Section \ref{section3}. {\figurename} \ref{figure4} plots the SOP versus $N$. As shown, the security can get improved via a larger array size, which consists with our former discussions. By \cite{b8}, $K$ in the FTR fading model denotes the power ratio between the dominant and remaining diffuse multipath components, and a larger value of $K$ yields a better channel condition. Hence, the SOP can get improved for a larger value of $K_{\rm B}$ or a smaller value of $K_{\rm E}$, which consists with the results in {\figurename} \ref{fig4a}. Besides, a smaller value of $m$ corresponds to heavier channel fluctuations and yields a worse channel condition \cite{b8}. Thus, the SOP can get improved for a small value of $m_{\rm E}$, which is verified by {\figurename} \ref{fig4b}.

\section{Conclusion}
\label{section4}
Novel expressions for the ASR and the SOP are derived to evaluate the secrecy performance of the FDA-assisted wiretap channels adopting finite-alphabet inputs. This work developed a generalized framework for secrecy performance analysis in wiretap channels driven by finite-alphabet signals. Based on the derivations, the properties of the FDA were discussed to offer important system design guides. 

\appendix[Proof of Lemma \ref{Lemma1}]
\label{Append1}
\begin{IEEEproof}
When $R_{s}+I_M\left(\gamma_{\rm E}\right)>\log_2 M$, we have $I_M\left(\gamma_{\rm B}\right)-I_M\left(\gamma_{\rm E}\right)<I_M\left(\gamma_{\rm B}\right)-\log_2 M+R_{s}<R_{s}$, which suggests that ${\mathcal I}_{s}<R_{s}$ always holds when $\gamma_{\rm B}>\gamma_{\rm E}$ and $R_{s}+I_M\left(\gamma_{\rm E}\right)>\log_2 M$. Moreover, when $R_{s}+I_M\left(\gamma_{\rm E}\right)\leq\log_2 M$, we find that $I_M\left(\gamma_{\rm B}\right)-I_M\left(\gamma_{\rm E}\right)<R_{s}$ is equivalent to ${I_M^{-1}\left(R_{s}+I_M\left(\gamma_{\rm E}\right)\right)}>\gamma_{\rm B}$. As a result, we obtain $\Pr\left({\mathcal I}_{s}<R_{s},\gamma_{\rm B}>\gamma_{\rm E}\right)=W_1+W_2$, where $W_1=\int_{0}^{\mathcal H}\int_{\gamma_{\rm E}}^{\phi_M\left(\gamma_{\rm E}\right)}f_{\rm{B}}\left(\gamma_{\rm{B}}\right)f_{\rm{E}}\left(\gamma_{\rm{E}}\right){\rm{d}}\gamma_{\rm{B}}{\rm{d}}\gamma_{\rm{E}}$,
$W_2=\int_{\mathcal H}^{+\infty}\int_{\gamma_{\rm E}}^{+\infty}f_{\rm{B}}\left(\gamma_{\rm{B}}\right)f_{\rm{E}}\left(\gamma_{\rm{E}}\right){\rm{d}}\gamma_{\rm{B}}{\rm{d}}\gamma_{\rm{E}}$, ${\mathcal H}\triangleq I_M^{-1}\left(\log_2 M-R_{s}\right)$ and $\phi_M\left(\gamma\right)\triangleq{I_M^{-1}\left(R_{s}+I_M\left(\gamma\right)\right)}$. Besides, $\Pr\left(\gamma_{\rm B}<\gamma_{\rm E}\right)$ is given by $\Pr\left(\gamma_{\rm B}<\gamma_{\rm E}\right)=Q_1+Q_2$, where $Q_1=\int_{0}^{\mathcal H}\int_{0}^{\gamma_{\rm E}}f_{\rm{B}}\left(\gamma_{\rm{B}}\right)f_{\rm{E}}\left(\gamma_{\rm{E}}\right){\rm{d}}\gamma_{\rm{B}}{\rm{d}}\gamma_{\rm{E}}$ and $Q_2=\int_{\mathcal H}^{+\infty}\int_{0}^{\gamma_{\rm E}}f_{\rm{B}}\left(\gamma_{\rm{B}}\right)f_{\rm{E}}\left(\gamma_{\rm{E}}\right){\rm{d}}\gamma_{\rm{B}}{\rm{d}}\gamma_{\rm{E}}$. Note that we can obtain $W_2+Q_2=\int_{\mathcal H}^{+\infty}f_{\rm{E}}\left(\gamma_{\rm{E}}\right)\int_{0}^{+\infty}f_{\rm{B}}\left(\gamma_{\rm{B}}\right){\rm{d}}\gamma_{\rm{B}}{\rm{d}}\gamma_{\rm{E}}=1-F_{\rm{E}}\left(\mathcal H\right)$ and $W_1+Q_1=\int_{0}^{\mathcal H}f_{\rm{E}}\left(\gamma_{\rm{E}}\right)\int_{0}^{\phi_M\left(\gamma_{\rm E}\right)}f_{\rm{B}}\left(\gamma_{\rm{B}}\right){\rm{d}}\gamma_{\rm{B}}{\rm{d}}\gamma_{\rm{E}}=\int_{0}^{\mathcal H}F_{\rm{B}}\left({\phi_M\left(\gamma_{\rm E}\right)}\right)f_{\rm{E}}\left(\gamma_{\rm{E}}\right){\rm{d}}\gamma_{\rm{E}}$. Based on \eqref{EQ19}, the SOP can be written as $P_{\rm{o}}\left(R_{s}\right)=\left(W_1+W_2\right)+\left(Q_1+Q_2\right)=\left(W_1+Q_1\right)+\left(W_2+Q_2\right)=\int_{0}^{\mathcal H}F_{\rm{B}}\left({\phi_M\left(\gamma_{\rm E}\right)}\right)f_{\rm{E}}\left(\gamma_{\rm{E}}\right){\rm{d}}\gamma_{\rm{E}}+1-F_{\rm{E}}\left(\mathcal H\right)$, which completes the proof of Lemma \ref{Lemma1}.
\end{IEEEproof}

\end{document}